%% file: OP-MMIMO.tex
\documentclass[conference]{IEEEtran}  

\usepackage{graphicx}
\usepackage{amssymb}
\usepackage{amsfonts}
\usepackage{latexsym}
\usepackage[cmex10]{amsmath}
\usepackage[usenames]{color}
\usepackage{array}
\usepackage{hyperref}
\usepackage{cite} 
\usepackage{import}   
\usepackage{booktabs} 

\def\mathlette#1#2{{\mathchoice{\mbox{#1$\displaystyle #2$}}%
                           {\mbox{#1$\textstyle #2$}}%
                           {\mbox{#1$\scriptstyle #2$}}%
                           {\mbox{#1$\scriptscriptstyle #2$}}}}
\renewcommand{\Vec}[1]{\mathlette{\boldmath}{#1}} 

\newcommand{\be}{\begin{equation}}
\newcommand{\ee}{\end{equation}}
\newcommand{\ba}{\begin{array}}
\newcommand{\ea}{\end{array}}
\newcommand{\bdm}{\begin{displaymath}}
\newcommand{\edm}{\end{displaymath}}
\newcommand{\bea}{\begin{eqnarray}}
\newcommand{\eea}{\end{eqnarray}}
\newcommand{\bean}{\begin{eqnarray*}}
\newcommand{\eean}{\end{eqnarray*}}

\def\argmin{\mathop{\text{argmin}}}

\def\diag{\text{diag}}

\def\E{\mathop{{}\mathbb{E}}}
\def\vects{\text{vec}}

\def\oH{\ensuremath{^\text{H}}} 
\def\oT{\ensuremath{^\text{T}}} 

\def\nfC{\ensuremath{f_\text{C}}}        

\def\nEb{\ensuremath{E_\text{b}}}        

\IEEEoverridecommandlockouts

\title{Combating Massive MIMO Channel Aging by Orthogonal Precoding}

\author{
\IEEEauthorblockN{Thomas Zemen and David L\"oschenbrand}
\IEEEauthorblockA{\textit{Security \& Communication Technologies, Center for Digital Safety \& Security}}
\textit{AIT Austrian Institute of Technology}\\
Vienna, Austria\\
\{thomas.zemen, david.loeschenbrand\}@ait.ac.at}

\begin{document}

\maketitle 

\begin{abstract}
In this work we investigate ultra-reliable low-latency massive multiple-input multiple-output (MIMO) communication links in vehicular scenarios, where coherence between uplink and downlink cannot be assumed. In such scenarios the channel state information obtained in the uplink will be outdated for the following downlink phase. To compensate for this channel aging we will utilize orthogonal precoding with two-dimensional precoding sequences in the time-frequency domain within an orthogonal frequency division multiplexing system. The channel hardening effect of massive MIMO transmission decreases, due to channel aging, with increasing frame duration and increasing velocity, while the channel hardening effect of orthogonal precoding (OP) increases with increasing time- and frequency-selectivity of the wireless communication channel. By combining massive MIMO and OP we can show by numeric link level simulation that the performance with outdated channel state information in terms of bit-error rate versus signal-to-noise ratio can be improved by two orders of magnitude.
\end{abstract}



\section{Introduction}
\label{sec:intro}
Ultra-reliable low-latency wireless communication (URLLC) links are an important component for connected autonomous vehicles, industrial wireless control loops, and many other machine-to-machine communication applications. The random fading process in wireless communication channels leads to signal strength fluctuations at the receive antenna and random unpredictable frame errors. 

Massive multiple-input multiple-output (MIMO) systems reach the capacity of multi-user MIMO systems by linear beam-forming over a large number of transmit antenna elements at the base station side \cite{Marzetta10}, achieving spatial channel hardening \cite{Gunnarsson18,Ngo17}. Beam-forming requires channel state information (CSI) at the transmitter side, which is obtain during a preceding uplink phase by exploiting channel reciprocity in a time-division duplex (TDD) system. 

For mobile users the channel impulse response is time-varying, hence the CSI becomes outdated (channel aging) due to the time delay between uplink and downlink transmission. Due to this outdated CSI the channel hardening effect of massive MIMO decreases with longer frame duration and increasing user velocity \cite{Truong13}. Previous work either considers a quasi static scenario where the uplink and downlink phase take part within a so called coherence interval \cite{Marzetta10} or performs channel prediction between the uplink and downlink transmission \cite{Truong13, Papazafeiropoulos15} using long-term statistical information.

Another method to improve the communication link reliability is orthogonal precoding (OP) \cite{Zemen18a,Hadani17,Hadani18}. OP spreads a data symbol in the time- and frequency domain and thus achieves also channel hardening, i.e. the fading variation of the received signal strength can be strongly reduced \cite{Zemen18a}. The channel hardening effect of OP increases with increasing time- and frequency seletivity (larger delay and Doppler spread) of the communication channel, see \cite[Table I]{Zemen18a}. Furthermore, Zemen et al. show in \cite{Zemen18a} that the channel hardening effect of OP is largely independent of the used set of orthogonal spreading sequences. All orthogonal constant modulus (CM) sequences forming unitary transforms, e.g., the discrete symplectic Fourier transform (DSFT) \cite{Hadani17} or the Walsh Hadamard transform (WHT) \cite{Shanks69}, achieve the same performance with respect to channel hardening \cite{Zemen18a}.

\subsection*{Contributions of the Paper:} 
\begin{itemize}
\item We provide a generalized channel hardening definition for massive MIMO with OP.
\item We show that the combination of massive MIMO with OP allows an efficient compensation of reduced spatial channel hardening by increased time-frequency channel hardening.
\end{itemize} 

\subsection*{Notation:}
We denote a scalar by $a$, a column vector by $\Vec{a}$ and its $i$-th element with $a_i$. Similarly, we denote a matrix by $\Vec{A}$ and its $(i,\ell)$-th element by $a_{i,\ell}$. The transpose of $\Vec{A}$ is given by $\Vec{A}\oT$ and its conjugate transpose by $\Vec{A}\oH$. A diagonal matrix with elements $a_i$ is written as $\diag(\Vec{a})$ and the $Q\times Q$ identity matrix as $\Vec{I}_Q$, respectivel y. The absolute value of $a$ is denoted by $\left|a\right|$ and its complex conjugate by $a^*$. The number of elements of set $\mathcal{I}$ is denoted by $|\mathcal{I}|$. We denote the set of all complex numbers by $\mathbb{C}$. The all one (zero) column vector with $Q$ elements is denoted by $\Vec{1}_Q$ ($\Vec{0}_Q$). We identify the 2D sequence $(a_{i,\ell})\in\mathbb{C}^{N \times M}$ for $i\in\{0,\ldots, N-1\}, \ell\in\{0,\ldots, M-1\}$ with the matrix $\Vec{A}\in\mathbb{C}^{N \times M}$, i.e., $\Vec{A}=(a_{i,\ell})$. Furthermore, we define the notation $\Vec{a}=\vects(\Vec{A})=\vects\big((a_{i,\ell})\big)\in\mathbb{C}^{MN \times 1}$, where $\vects(\Vec{A})$ denotes the vectorized version of matrix $\Vec{A}$, formed by stacking the columns of $\Vec{A}$ into a single column vector.


\section{Signal Model for OP in Massive MIMO Systems}
\label{se:SignalModel}

In this work we are concerned with URLLC links for highly mobile users. Hence, the typical packet duration is short and the required reliability of the communication link shall be as high as possible. Due to short packet length the diversity utilized by the channel code is limited. Hence, additional linear precoding methods are crucial to utilize the full channel diversity in time, frequency and space, enabling URLLC.

We combine two linear preprocessing techniques in this work:
\begin{itemize}
\item The first one is OP, which exploits diversity in the time- and frequency domain, and is applied once for each data packet. OP achieves channel hardening by appropriate precoding at the transmitter side and parallel interference cancellation (PIC) at the receiver side \cite{Zemen18a}. The channel hardening effect of OP increases with the delay- and Doppler spread of the doubly selective fading process, as well as with increasing extent of the precoding region in time- and frequency \cite{Zemen18a}. 
\item The second preprocessing technique is maximum-ratio beam-forming in a massive MIMO system. Beam-forming uses weights that are specific for each antenna element at the transmitter side. It achieves channel hardening that increases with the number of transmit antennas but decreases with (a) increasing frame duration and (b) increasing velocity of the mobile station, due to channel aging.
\end{itemize}
Throughout the paper, we will use the term \emph{precoding} to describe linear operations performed in the time-frequency domain and the term \emph{beam-forming} for the linear operations in the spatial domain.

\subsection{Precoding}
We precode data symbols $b_{p,n}\in\mathcal{A}$, $p\in\{0,\ldots, N-1\}$, $n\in\{0,\ldots, M-1\}$ on a general transform domain grid, with a general complete set of 2D orthonormal basis functions:
\be
d_{q,m}=\sum_{p=0}^{N-1} \sum_{n=0}^{M-1} b_{p,n} s^{p,n}_{q,m}\,, 
\label{eq:Precoding}
\ee
where $s^{p,n}_{q,m}$ denotes two-dimensional precoding sequences and $d_{q,m}$ the result of the precoding operation, respectively. The time-frequency grid is defined by the discrete time index $m\in\{0,\ldots, M-1\}$ and the discrete frequency index $q\in\{0,\ldots, N-1\}$. The quadrature amplitude modulation alphabet is denoted by $\mathcal{A}$. 

Let $\Vec{B}\in\mathcal{A}^{N \times M}$ denote the symbol matrix with elements $b_{p,n}$. We define the symbol vector $\Vec{b}=\vects(\Vec{B})=\vects\big((b_{p,n})\big)\in\mathcal{A}^{MN \times 1}$, and the precoded symbol vector $\Vec{d}=\vects\big((d_{q,m})\big)$, using the notation introduced in Sec.~\ref{sec:intro}. Matrix $\Vec{S}=[\Vec{s}_{0,0}, \ldots, \Vec{s}_{N-1,0}, \Vec{s}_{0,1}, \ldots, \Vec{s}_{N-1,M-1}]\oT\in\mathbb{C}^{MN\times MN}$
collects all vectorized 2D precoding sequences $\Vec{s}_{p,n}=\vects\big((s_{q,m}^{p,n})\big)$ 
column-wise. With these definitions, we can write \eqref{eq:Precoding} in vector matrix notation as $\Vec{d}=\Vec{S}\Vec{b}$.

\subsection{Massive MIMO Beam-Forming}
The precoded data symbol vector $\Vec{d}$ is transmitted from all $A$ antenna elements of the massive MIMO base station after linear beam-forming. We extend the signal model introduced in \cite{Zemen18a} for the massive MIMO case. The received samples at the single antenna of the mobile station
\be
\Vec{\psi} =\Vec{G}\oT \Vec{\Omega} \Vec{d} + \frac{1}{\sqrt{\rho}}\Vec{n}\,,\label{eq:OPMMIMODL}
\ee
where the massive MIMO channel matrix
\be
\Vec{G}=\left[
\begin{matrix}
\diag(\Vec{g}_{1})\\
\vdots \\
\diag(\Vec{g}_{A}) \\ 
\end{matrix}
\right]\,.
\label{eq:Gmmimo}
\ee
The time-variant frequency response from antenna $a$ to the mobile station is denoted by $\Vec{g}_{a}=\vects\big((g_{q,m}^{a})\big)$. Vector $\Vec{g}_a$, $a\in\{1, \ldots, A\}$ represents the combined result of OFDM modulation, the doubly selective channel, and OFDM demodulation between base station antenna $a$ and the mobile station. The number of antenna elements is denoted by $A$. We use the normalization $\E\{||\Vec{G}/\sqrt{N}||^2_2\}=M$ such that the array gain is not taken into account. Additive white complex symmetric Gaussian noise is denoted by $\Vec{n}$ with zero mean and variance $\Vec{I}_{AMN}$, $\Vec{n}=\mathcal{CN}(0,\Vec{I}_{AMN})$. Linear massive MIMO beam-forming is performed by $\Vec{\Omega}$, the signal to noise ratio (SNR) at the receiver side is denoted by $\rho$. The beam-forming matrix
\be
\Vec{\Omega}=
\left[
\begin{matrix}
\diag\left(\vects\left((\omega_{q,m}^1)\right)\right)\\
\vdots \\
\diag\left(\vects\left((\omega_{q,m}^A)\right)\right)\\ 
\end{matrix}
\right]
\ee
has the same structure as $\Vec{G}$ in \eqref{eq:Gmmimo}.

We specifically analyze maximum ratio transmission with
\be
\Vec{\Omega}=\frac{{\Vec{\tilde{G}}}^*}{||\Vec{\tilde{G}}||}\,,
\label{eq:MRbeamforming}
\ee
where the channel estimates at the base station side $\Vec{\tilde{G}}$ are assumed according to a general error model
\be
\Vec{\tilde{G}}=\Vec{G}+\Vec{E}\,.
\label{eq:ErrorModel}
\ee

The combined channel between transmitter and receiver including the massive MIMO beam-forming weights results in a diagonal matrix
\be
\diag(\Vec{\phi})=\Vec{G}\oT\Vec{\Omega}
\label{eq:EffChannel}
\ee
with
\be
\Vec{\phi}=\vects\big((\phi_{q,m})\big)= \frac{1}{||\Vec{G}||}\vects\Big(\big(\sum_{a}g_{q,m}^a {\left.\tilde{g}_{q,m}^a\right.}^*\big)\Big)\,.
\ee
In the case of $\Vec{E}=\Vec{0}$ it simplifies to
\be
\Vec{\phi}=\frac{1}{||\Vec{G}||}\vects\Big(\big(\sum_{a}{|g_{q,m}^a}|^2\big)\Big)\,,
\ee
i.e., for each element $(q,m)$ of the time-frequency grid, maximum ratio combining is achieved. With beam-forming according to \eqref{eq:MRbeamforming}, the combined channel becomes almost frequency flat and non time-selective (assuming favorable propagation conditions). For $A\rightarrow\infty$ these conclusion becomes exact \cite{Marzetta10}.

Inserting \eqref{eq:EffChannel} into \eqref{eq:OPMMIMODL} we obtain
\be
\Vec{\psi} = \diag(\Vec{\phi}) \Vec{S}\Vec{b} + \frac{1}{\sqrt{\rho}}\Vec{n}\,.
\ee
We defined the effective spreading sequence as $\Vec{\tilde{S}}=\diag(\Vec{\phi}) \Vec{S}$, resulting in
\be
\Vec{\psi}= \Vec{\tilde{S}}\Vec{b} + \frac{1}{\sqrt{\rho}}\Vec{n}\,.
\label{eq:OPMMIMODLeff}
\ee 

\subsection{Iterative Detection for Massive MIMO Systems} 
\label{sec:Detection}
In this section we briefly revisit the iterative PIC algorithm \cite{Zemen06c}, that was introduced in \cite{Zemen18a} for OP. In the first iteration, $i=1$, data symbol estimates 
\be
\hat{\Vec{b}}=  \Vec{S}\oH \Vec{W} \Vec{\psi}= \Vec{S}\oH \Vec{W} \Vec{\tilde{S}} \Vec{b} + \frac{1}{\sqrt{\rho}}\Vec{n}
\label{eq:ItOne}
\ee
are obtained by weighting (windowing) with 
\be
\Vec{W}=\diag\left(\vects\left(\frac{\phi_{q,m}^*}{|\phi_{q,m}|^2+1/\rho}\right)\right)
\ee
according to a minimum mean square error (MMSE) criterion and matched filtering with $\Vec{S}\oH$. 

We proceed with PIC for all following iterations $i>1$: Soft-symbols from iteration $i-1$ are used to express PIC for grid element $(p,n)$ according to
\begin{subequations}
\begin{align}
\alpha_{p,n}^{(i)}&=\Vec{\tilde{s}}_{p,n}\oH\left(\Vec{\psi}-\tilde{\Vec{S}}\tilde{\Vec{b}}^{(i-1)} + \tilde{\Vec{s}}_{p,n}\tilde{b}_{p,n}^{(i-1)}\right)\nonumber\\ \label{eq:PIC}
&\approx \underbrace{\Vec{\tilde{s}}_{p,n}\oH \tilde{\Vec{s}}_{p,n}}_{\gamma_{p,n}} b_{p,n} + \tilde{\Vec{s}}_{p,n}\oH \frac{1}{\sqrt{\rho}}\Vec{n}\\ 
&=\gamma_{p,n} b_{p,n} + \frac{1}{\sqrt{\rho}}\tilde{n}_{p,n}\,,
\label{eq:effSignalModel}
\end{align}
\end{subequations}
where the soft-symbol feedback vector $\tilde{\Vec{b}}^{(i)}=\vects\big((\tilde{b}_{p,n}^{(i)})\big)$ and the effective channel coefficient is denoted by $\gamma_{p,n}$. Noise $\tilde{n}_{p,n}$ has the same distribution as $n_{q,m}$. 

The symbol-wise ML expression 
\be 
\hat{b}_{p,n}=\argmin_{b_{p,n} \in\mathcal{A}}\{| \alpha_{p,n} - \gamma_{p,n} b_{p,n} |^2\}
\label{eq:TFML}
\ee
for data symbol $b_{p,n}$ is formulated based on the scalar signal model \eqref{eq:effSignalModel}. A soft-output sphere decoder \cite{Studer08}, using \eqref{eq:TFML}, supplies log-likelihood ratios (LLRs) $L_k$. The LLRs are used as input for the BCJR decoder \cite{BCJR74a}.


%
%
%
%
%
%
%
%
%
%
%
%
%

\section{Channel Hardening in a massive MIMO system with OP}

In \cite{Fatema18}, massive MIMO beam-forming methods are compared, assuming accurate CSI is available at the base-station side. In URLLC applications for highly mobile users this assumption is hard to maintain. For time-variant scenarios, we investigate the joint usage of a massive MIMO system with OP aiming to minimize the bit error rate (BER). 
	
The joint analysis of a combined system with massive MIMO and OP is obtained by utilizing the effective channel coefficient $\gamma_{p,n}$ in \eqref{eq:effSignalModel}. Generalizing the results of \cite{Zemen18a} we obtain for CM precoding sequences
	\be
	\gamma_{p,n}^\text{CM}=\gamma=\frac{1}{MN}\sum_q\sum_m\left|\sum_a \frac{g_{q,m}^a {\left. {{\tilde{g}_{q,m}}^{a}}\right.}^*}{||\Vec{\tilde{G}}||}\right|^2\,.
	\ee
Please note that all grid elements $\gamma_{p,n}^\text{CM}=\gamma$ will be effected by the same effective channel condition and that we omit the superscript in the following.
	
We can evaluate the channel hardening effect in the combined OP massive MIMO system empirically by analyzing the distribution of $\gamma$ for $F$ frame transmissions \cite{Gunnarsson18,Ngo17}: 
	\be
	\mu_\gamma = \frac{1}{F} \sum_{f=1}^{F} \gamma[f]\,,
	\ee
	
	\be
	\sigma_\gamma= \sqrt {\frac{1}{F} \sum_{f=1}^{F} \left(\gamma[f] - \mu_\gamma\right)^2}\,.
	\ee
	Here $\mu_\gamma$ denotes the mean of $\gamma$ (first moment) and $\sigma_\gamma$ the standard deviation (root of the second central moment). Channel hardening is measured as
	\be
	\beta=\frac{\sigma_\gamma}{\mu_\gamma}\,,
 	\ee
following the definition in \cite{Gunnarsson18,Ngo17}.	
The value of $\beta\rightarrow 0$ for $A,M,N\rightarrow \infty$.

%
%
	
We expect, that increasing $||\Vec{E}||$ in \eqref{eq:ErrorModel} will reduce the channel hardening effect of massive MIMO and the combined channel $\Vec{\phi}$ will become time- and frequency selective. Hence, we explore the benefit of OP for different error models representing two extreme cases:
	\begin{enumerate}
		\item \emph{Perfect CSI beam-forming (PERBF):} We assume the base station knows the time-variant CSI for the downlink phase perfectly, i.e. $\tilde{g}_{q,m}^a= g_{q,m}^a$ and $\tilde{e}_{q,m}^a= 0$\,.
		
		\item \emph{Block fading beam-forming (BFBF):} The base station uses the last known CSI from the end of the uplink transmission for precoding during the full downlink frame: $\tilde{g}_{q,m}^a=g_{q,-1}^a \quad\forall\quad m\in\{0,\ldots, M-1\}\,$, resulting in $e_{q,m}^a=g_{q,m}^a - g_{q,-1}^a\,$.
		
		
		
	\end{enumerate}
	
	%
	
	

%
	
	The case of massive MIMO without OP (NO) can be analyzed in the same frame work by setting the precoding matrix $\Vec{S}=\Vec{I}_{MN}$ which leads to $\gamma^{(\text{NO})}_{p,n}=|\phi_{q,m}|^2$ with $p=q$ and $n=m$, see \cite{Zemen18a}.

\section{Numerical Simulations Results}
We provide numerical simulation results for an orthogonal frequency division mutiplexing (OFDM) system with parameters similar to IEEE 802.11p. The bandwidth $B=10\,\text{MHz}$, the number of subcarriers $N=64$, the cyclic prefix has length of $G=16$ samples and the frame length $M=44$ OFDM symbols. The carrier frequency is $\nfC=5.9\,\text{GHz}$. We use a geometry based channel model (GCM) with an exponential decaying power delay profile with a root mean square delay spread of $0.4\,\mu\text{s}$ and a Clarke Doppler spectrum \cite{Clarke68} for each channel tap, see \cite{Zemen18a}. The mobile station has a single antenna and moves with $v=200\,\text{km/h}$.


%

\subsection{Combined Channel}
\begin{figure} 
	\centering
	\includegraphics[width=\columnwidth]{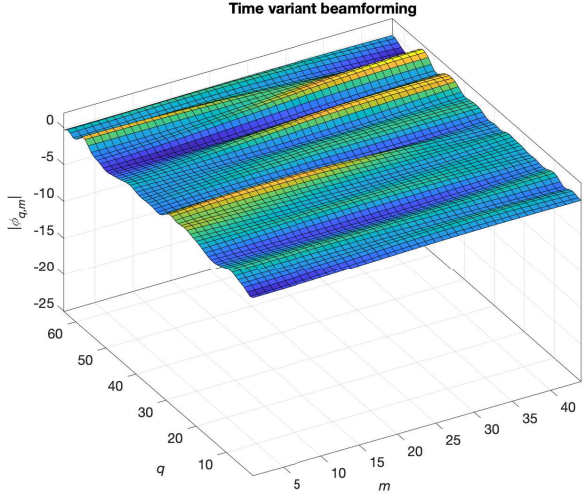}
	\caption{Absolute value of the combined channel $|\phi_{q,m}|$ for time-variant massive MIMO beam-forming with perfect CSI (PERBF) using $A=64$ antennas.}
	\label{fig:CombPERBF}
\end{figure}

In Fig. \ref{fig:CombPERBF} 
we plot the combined channel $|\phi_{q,m}|$, versus time $m$ and frequency $q$. Due to maximum ratio beam-forming using a perfectly known channel (PERBF) we achieve a nearly frequency flat and non time-selective frequency response.

\begin{figure} 
	\centering
	\includegraphics[width=\columnwidth]{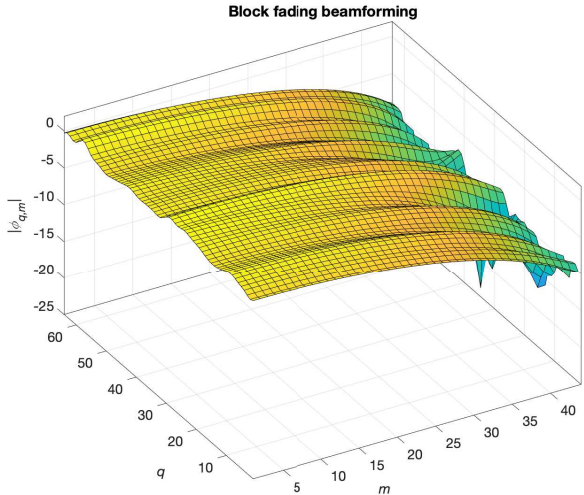}
	\caption{Absolute value of the combined channel $|\phi_{q,m}|$ for block-fading massive MIMO beam-forming (BFBF) using $A=64$ antennas.}
	\label{fig:CombBFBF}
\end{figure}

In Fig. \ref{fig:CombBFBF} 
we show the same plot, but now the last CSI during uplink transmission is used for the full duration of the downlink frame, we term this approach block-fading beam-forming (BFBF). The channel aging effect is demonstrated, i.e. channel hardening decreases with increasing time $m$.

\subsection{Channel Hardening}
\begin{figure} 
	\centering
	\includegraphics[width=\columnwidth]{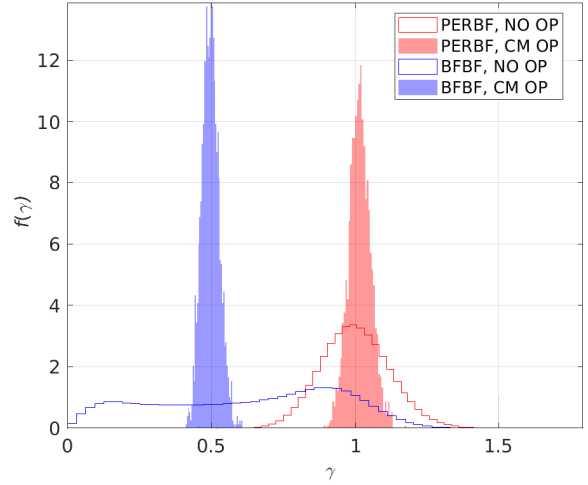} 
	\caption{Pdf $f(\gamma)$ using block-fading beam-forming (BFBF) or time-variant beam-forming with perfect CSI (PERBF) either with constant modulus orthogonal precoding (CM OP) or without precoding (NO OP).} 
\label{fig:EffChHist}
\end{figure}

In Fig. \ref{fig:EffChHist}
we plot the empirical probability density function (pdf) $f(\gamma)$ of the effective channel coefficient for $F=2000$ frames. 
The red pdf depicts the distribution for a massive MIMO system with perfect beam-forming (PERBF) and with NO OP. The solid red pdf shows the histogram for a massive MIMO with PERBF and with CM OP. Channel hardening increases by combining PERBF with OP. 

Massive-MIMO beam-forming using a constant channel state (BFBF) and with NO OP leads to an effective channel that has a large standard deviation as depicted by the blue pdf in Fig. \ref{fig:EffChHist}. Channel hardening is strongly increased by using OP together with BFBF, i.e. the standard deviation decreases substantially, see the solid blue pdf. The average received power (represented by the mean of the pdf, $\mu_\gamma$) decreases due to the non-constructive superposition of transmit signals for $m>0$. This effect can only be reversed by using channel prediction at the base station.

In Table \ref{tab:ChannelHardening}
\begin{table}
\begin{center}
\caption{Estimated mean and standard deviation comparing for the effective channel coefficient $\gamma$.}
\label{tab:ChannelHardening}
\begin{tabular}{ll} 
\toprule
Type	      				& $\beta$\\
\midrule
PERBF + NO PC		& 	 $0.12$ \\
PERBF + CM PC		& 	 $0.04$ \\
BFBF + NO PC		& 	 $0.50$ \\
BFBF + CM PC		& 	 $0.06$ \\
\bottomrule
\end{tabular}
\end{center}
\end{table}
we show the empirical results for (a) beam-forming with perfect CSI (PERBF) and  (b) block fading beam-forming using the last channel state from the preceding uplink transmission (BFBF).  For both cases we show the results for using either only massive MIMO beam-forming with $A=64$ antennas or massive MIMO beam-forming and constant modulus (CM) OP.

\subsection{Bit Error Rate} 
\begin{figure} 
	\centering
	\includegraphics[width=\columnwidth]{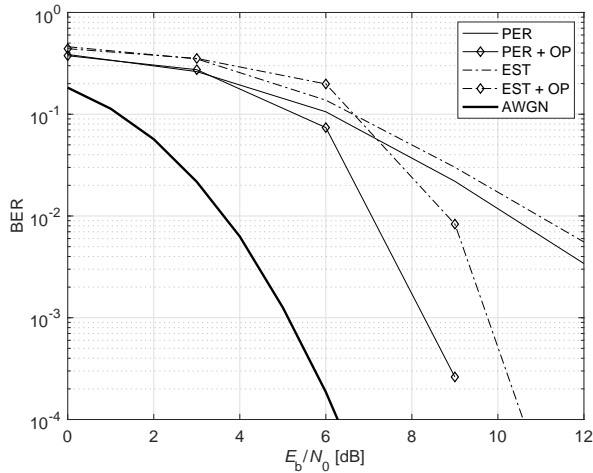}
	\caption{BER versus $\nEb/N_0$ comparing massive MIMO with $A=64$ antennas with and without OP. Furthermore we show results for perfect CSI (PER) and for channel estimation (EST) for a single user with velocity of $v=200\,\text{km/h}$. The iterative massive MIMO OP receiver uses four iterations \cite[Fig. 1]{Zemen18a}.}
	\label{fig:OPDSFT}
\end{figure}

In Fig. \ref{fig:OPDSFT} 
the BER vs. SNR of a massive MIMO link to a single vehicle moving with 200km/h is shown using numerical link level simulation results. All simulation results are calculated for the case of precoding at the base station using the last know CSI from the preceding uplink transmission (BFBF).

The full line shows the performance with perfectly (PER) known CSI at the receiver side.  Due to the time-variance of the communication channel the channel hardening effect at the transmitter is lost and the BER performance is substantially reduced. By adding OP the lost channel hardening can be re-obtained, this case is shown by the full line with diamonds (PER+OP). A similar behavior is also obtained in the case of using channel estimates (EST; EST+OP) at the receiver side. The AWGN curve shows the best case performance of the used convolutional code in a pure additive white Gaussian noise channel without fading. Fig. \ref{fig:OPDSFT} demonstrates that by combining massive MIMO with OP the BER performance can be improved by more than two orders of magnitude. 

\section{Conclusions} 
\label{sec:Conclusions}
In this paper we have investigated a new approach to combat channel aging in massive MIMO systems that operates in highly time-variant channel conditions. In such scenarios the channel state information obtained in the uplink will be outdated for the following downlink phase causing strongly reduced channel hardening. We showed for the first time that channel aging can be compensated by orthogonal precoding with two-dimensional precoding sequences in the time-frequency domain. We introduced a generalized channel hardening definition for massive MIMO with OP. Using this metric we showed that the combination of massive MIMO with OP allows an efficient compensation of reduced spatial channel hardening by increased time-frequency channel hardening. This result is validated by link level simulation results in terms of BER vs. $\nEb/N_0$. The combination of massive MIMO and OP improves the BER by more than two orders of magnitude in highly time-variant scenarios.

\section*{Acknowledgments} 
This work is funded by the Austrian Research Promotion Agency (FFG) and the Austrian Ministry for Transport, Innovation and Technology (bmvit) within the project MARCONI (861208) of the funding program ICT of the Future.



\input{OP-MMIMO.bbl}

\end{document}